\documentclass[prx,twocolumn,superscriptaddress,nopacs,amsmath,amssymb]{revtex4-2}
\usepackage{graphicx}
\graphicspath{{./}{./figs/}}

\usepackage{hyperref}

\usepackage{booktabs}
\usepackage{makecell}

\newcommand{\Var}{\mathrm{Var}}

\begin{document}

\title{Stability and Absence of a Tower of States in Ferrimagnets}
\author{Louk Rademaker}
\affiliation{Department of Theoretical Physics, University of Geneva, 1211 Geneva, Switzerland}
\author{Aron Beekman}
\affiliation{Department of Physics, Keio University, 3-14-1 Hiyoshi, Kohoku-ku, Yokohama 223-8522, Japan}
\author{Jasper van Wezel}
\affiliation{Institute for Theoretical Physics Amsterdam, University of Amsterdam, Science Park 904, 1098 XH Amsterdam, The Netherlands}

\date{\today}

\begin{abstract}
Antiferromagnets and ferromagnets are archetypes of the two distinct (type-A and type-B) ways of spontaneously breaking a continuous symmetry. Although type-B Nambu--Goldstone modes arise in various systems, the ferromagnet was considered pathological due to the stability and symmetry-breaking nature of its exact ground state. However, here we show that symmetry-breaking in ferrimagnets closely resembles the ferromagnet. In particular, there is an extensive ground state degeneracy, there is no Anderson tower of states, and the maximally polarized ground state is thermodynamically stable. Our results are derived analytically for the Lieb--Mattis ferrimagnet and numerically for the Heisenberg ferrimagnet. We argue that these properties are generic for type-B symmetry-broken systems, where the order parameter operator is a symmetry generator. 
\end{abstract}

\maketitle

\section{Introduction}
Spontaneous symmetry breaking (SSB) is the phenomenon that the thermal equilibrium state of a many-body system has lower symmetry than the Hamiltonian that governs it. For a continuous symmetry, there is a multitude of degenerate symmetry-breaking {\em ground} states in the thermodynamic limit where the number of constituents $N$ goes to infinity. Conversely when $N$ is finite, most such systems possess a unique and symmetric ground state. This was explicitly shown for the Heisenberg antiferromagnet by Marshall, and Lieb and Mattis~\cite{1955RSPSA.232...48M,Lieb:1962hn}, and is now understood to be quite general~\cite{Koma:1994,Tasaki:2019}. However, these unique ground states are not stable, in the sense that adding even a small symmetry-breaking perturbation $\epsilon$ will lead to a symmetry-broken ground state; in the thermodynamic limit an infinitesimal perturbation suffices to break the symmetry. Examples of this type of symmetry breaking include the breaking of $\mathbb{Z}_2$ (up--down) symmetry in Ising models, of $U(1)$ (phase rotation) symmetry in $XY$-models, of $SU(2)$ (spin-rotational) symmetry in Heisenberg antiferromagnets, and of translational symmetry breaking in crystals~\cite{BRvW:lecturenotes}.

The instability of the symmetric ground state of finite-sized systems may be intuitively understood by realizing it is actually a type of `Schr\"odinger cat state', namely a superposition of macroscopically distinct states which each break the symmetry differently~\cite{BRvW:lecturenotes,Louk:2019}. There must then be some observable $A$ with an extensive expectation value $\langle A \rangle \sim \mathcal{O}(N)$ whose variance scales as $\Var A \sim N^2$, violating the cluster decomposition property~\cite{Shimizu:2002JPSJ}. In other words, the symmetric ground state contains macroscopic uncertainty
of an extensive observable, making it exceedingly susceptible to local perturbations. The symmetry-broken state, on the other hand, does not contain macroscopic uncertainties and is thermodynamically stable. All the while it is not an energy eigenstate; instead, it is a superposition of the ground state and zero-wavenumber low-energy states. If the broken symmetry is continuous, the gap of these states is of the order $\mathcal{O}(1/N)$. The existence of this low-energy {\em tower of states}
was observed for quantum antiferromagnets by Anderson~\cite{Anderson:2011vu}, and subsequently shown to be generic in SSB systems with a symmetric ground state~\cite{Horsch:1988,Kaiser:1989,Kaplan:1990,1993PhRvL..70.2483A,1992PhRvL..68.1762A,BBernu:2011te,Koma:1994,Tasaki:2019,Kallin2011PhRvB..84p5134K,Metlitski:akSbyvLD,Rademaker:2015ie}. To understand the physics of finite-size systems with SSB, stability is at least as important as the energy spectrum~\cite{Shimizu:2002PRL}. The existence of the tower of states can be an important numerical diagnostic to show the propensity to SSB even in very small systems~\cite{Lhuillier:2005,Lauchli:2016}.

This behavior is however not completely general. Indeed it has long been known that the Heisenberg ferromagnet has degenerate, symmetry-breaking ground states for systems of any size. Furthermore its order parameter is a symmetry generator itself and hence a conserved quantity~\cite{Anderson:InL7mPjj}, there is only a single Nambu--Goldstone modes while two symmetry generators are broken, and this Goldstone mode has a quadratic dispersion. It has recently been cleared up that these two features go hand-in-hand: whenever the commutator of two broken symmetries has a non-vanishing expectation value, two Goldstone fields conspire to form a single, quadratically dispersing gapless mode accompanied by a gapped partner mode.\cite{Brauner:2010review,WatanabeBrauner:2011,Watanabe:2012jn,Hidaka:2013,Hayata:2015} Such Goldstone modes have been dubbed type-B, while ordinary, linearly dispersing Goldstone modes are called type-A.

It is now the question whether the other ferromagnet phenomenology---degenerate, thermodynamically stable finite-size ground states and no tower of states---also carries over to any type-B SSB system. A natural starting point is the {\em ferrimagnet}, a state with antiferromagnetic correlations between two unequal-size spin species, which implies in addition to antiferromagnetic order {\em also} ferromagnetic order. 
Earlier, one of the authors suggested that ferrimagnets would feature a tower of states, since their classical ground states are not eigenstates of the quantum Hamiltonian~\cite{Beekman:2015gr}. On the other hand, it has long been known that spin systems with any non-zero magnetization have macroscopically degenerate ground states~\cite{Lieb:1962hn,Tasaki:2019book}.

Here we show that the Heisenberg ferrimagnet is far more akin to a ferromagnet than to an antiferromagnet: we demonstrate explicitly that there exists a thermodynamically stable finite-size ground state, and that there is no tower of states separated from the ground state by an excitation gap of order $\mathcal{O}(1/N)$.  This stable ground state can be understood to be a classical (product) state supplemented by quantum corrections, in the same way that the SSB states of type-A systems are~\cite{Anderson:2011vu,Anderson:InL7mPjj}.
We provide an analytic derivation of the stability of this state in the simplified case of the Lieb--Mattis model, and provide numerical evidence for the stability in the full Heisenberg Hamiltonian.

We furthermore argue that this behavior is general for any system with exclusively type-B SSB.
This paints a comprehensive picture of SSB: if the ground state is unique, it must be accompanied by a tower of states in order for thermodynamically stable SSB states, as a superposition of very-closely spaced energy eigenstates, to exist. In type-B systems such a tower of states is absent, but SSB is possible because thermodynamically stable, symmetry-breaking exact ground states exist even for finite-size systems.

This article is organized as follows. In Section~\ref{sec:Antiferromagnets} we briefly outline the Lieb--Mattis argument which leads to the tower of states in antiferromagnets, the archetype for type-A SSB. In Section~\ref{SecToS} we show that a tower of states in absent in Heisenberg ferrimagnets, while there is a ground state degeneracy. These ground states all break the $SU(2)$ spin-rotation symmetry as is shown in Section~\ref{sec:Spontaneous symmetry breaking}. In Section~\ref{sec:Neel} we calculate the overlap of the SSB ground state with the classical N\'eel state and compare with the situation in the antiferromagnet. The central part of this work is the demonstration that the two ground states which have maximal positive or negative magnetization are thermodynamically stable. In Section~\ref{SecStab} this is shown analytically for the Lieb--Mattis model. Numerical evidence of the stabily of small 1D ferrimagnets is provided in Section~\ref{Sec:NumResults} by exact diagonalization. We conclude with a comprehensive picture of SSB and directions for further research in Section~\ref{sec:Outlook}.

\section{Antiferromagnets}\label{sec:Antiferromagnets}
To set the stage, we shall first recall some well-known facts about Heisenberg antiferromagnets on bipartite lattices. The Heisenberg Hamiltonian is
\begin{equation}\label{eq:Heisenberg Hamiltonian}
 	\hat{H}_{\mathrm{H}} = J\sum_{\langle ij \rangle} \hat{\vec{S}}_i \cdot \hat{\vec{S}}_j.
\end{equation}
Here $\hat{\vec{S}}_i$ is a spin-$s$ operator\footnote{We will denote all operators with a `hat' $\hat{~}$ to avoid confusion with their eigenvalues.} on site $i$, the sum is over nearest-neighbor lattice sites, and $J >0$ is a coupling constant. This Hamiltonian is invariant under global $SU(2)$-spin rotations.
If the lattice is bipartite it can be divided in $A$- and $B$-sublattices such that each site has neighbors only on the other sublattice.  The classical ground states are N\'eel states with spins anti-aligned on the two sublattices, breaking the $SU(2)$ symmetry to $U(1)$; the direction of N\'eel ordering is spontaneously chosen. If furthermore the number of sites of each sublattice is the same (for instance in square and hexagonal lattices), there is no net magnetization $\langle \sum_i  \hat{\vec{S}}_i \rangle = 0$.

The N\'eel states are not eigenstates of Eq.~\eqref{eq:Heisenberg Hamiltonian}, and will be affected by quantum corrections. But even stronger, a finite-size system governed by this Hamiltonian has a unique ground state with total spin value $S=0$, which therefore does not break any symmetry. This was shown by Marshall~\cite{1955RSPSA.232...48M}, and can be understood due to an elegant argument by Lieb and Mattis~\cite{Lieb:1962hn}: consider the following Hamiltonian (``Lieb--Mattis model'')
\begin{equation}\label{eq:Lieb--Mattis model}
	\hat{H}_\mathrm{LM}  
		= \frac{2J}{N} \hat{\vec{S}}_A \cdot \hat{\vec{S}}_B 
		= \frac{J}{N} (\hat{S}^2 - \hat{S}_A^2 - \hat{S}_B^2),
\end{equation}
where $\hat{\vec{S}} = \sum_i \hat{\vec{S}}_i$ is the total spin of the system, 
$\hat{\vec{S}}_{A,B} = \sum_{i \in A,B} \hat{\vec{S}}_i$ the total sublattice spin, 
 and $\hat{S}^2 = \hat{\vec{S}}^2$ etc. 
Note that $\hat{\vec{S}} = \hat{\vec{S}}_A + \hat{\vec{S}}_B$. 
In this model each spin on the $A$-sublattice interacts with all spins on the $B$-sublattice and vice versa. This Hamiltonian simultaneously commutes with $\hat{S}^2_A$, $\hat{S}^2_B$, $\hat{S}^2$ and $\hat{S}^z$, and eigenstates can therefore be designated by the quantum numbers $\lvert S_A S_B S M_z \rangle$, with energies $E = \frac{J}{N} \big( S(S+1) - S_A(S_A +1) - S_B(S_B+1) \big)$. Clearly the energy is minimal when $S$ is minimal and both $S_A$ and $S_B$ are maximal. The minimal value is $S=0$, and therefore the ground state is a total spin singlet, is unique, and does not break any symmetry. 

It can be easily seen that the Lieb--Mattis model is equal to only the $\vec{k}=\vec{0}$ and $\vec{k} = \vec{Q}  = (\pi,\pi,\ldots,\pi)$
contributions of the Fourier-transformed Heisenberg Hamiltonian Eq.~\eqref{eq:Heisenberg Hamiltonian}.
Lieb and Mattis have shown that for any finite $N$ the overlap between the ground state of Eq.~\eqref{eq:Lieb--Mattis model} and the ground state of Eq.~\eqref{eq:Heisenberg Hamiltonian} is non-vanishing. 
Therefore, these two states must have the same quantum numbers, and also the ground state of the Heisenberg antiferromagnet is a total spin singlet.

The correspondence between the two Hamiltonians goes further. Excitations that keep $S_A$ and $S_B$ fixed while increasing $S$
cost an energy $\mathcal{O}(J/N)$. For large $N$, these energy levels are almost degenerate with the ground state. There is therefore a tower of extremely low-lying states with $S_A$ and $S_B$ maximal, $M_z =0$, and differing $S$, and this carries over to the Heisenberg antiferromagnet by the same argument. 
Note that excitations in the Lieb-Mattis model that change $S_A$ or $S_B$ cost energy of at least $\mathcal{O}(J)$, just as local excitations such as spin flips, while Nambu--Goldstone modes (spin waves) in the Heisenberg model have lowest energy $\mathcal{O}(J/L)$ with $L$ the linear size of the system.

The variance of the local N\'eel order parameter $\hat{\mathcal{N}}^z_i = (-1)^i \hat{S}^z_i$ in the symmetric ground states is of order one~\cite{Tasaki:2019}, so that variance of the total N\'eel order parameter scales as $N^2$, indicating that this state is not thermodynamically stable. This is easy to see, when one realizes the ground state of the Lieb--Mattis model is equal to the equal-weight superposition of  classical N\'eel states in all magnetization directions~\cite{Louk:2019}. This is a Schr\"odinger cat state, which is extraordinarily sensitive to external perturbations.

\section{The Absence of a Tower of States}
\label{SecToS}
We will now begin demonstrating the differences with the picture painted in Section~\ref{sec:Antiferromagnets}, for the case where the magnetization is finite. For concreteness, we study the Heisenberg ferrimagnet governed by Hamiltonian Eq.~\eqref{eq:Heisenberg Hamiltonian}, but now the spins on $A$- and $B$-sublattices are $s_A$ and $s_B$ respectively, with $s_A \neq s_B$. Without loss of generality we choose $s_A>s_B$. On a bipartite lattice with equal number of sublattice sites, the classical N\'eel state has a finite magnetization $\langle \hat{S}^z \rangle = N(s_A - s_B)/2$, and staggered magnetization $\langle \hat{\mathcal{N}}^z \rangle =  N(s_A + s_B)/2$. Everything we say here also holds for antiferromagnets where the number of $A$-sublattice sites is different from number the $B$-sublattice sites; the imporant feature is that the total spin is $S =  \lvert S_A- S_B\rvert > 0$  where $S_{A,B} = \sum_{i \in A,B} s_{A,B}$.

It is known that the number of ground states of this model is equal to $2S +1 = 2\lvert S_A - S_B \rvert +1 = N\lvert s_A - s_B \rvert + 1$~\cite{Lieb:1962hn,Tasaki:2019book}, which can again be inferred from the overlap of these states with the ground states of the corresponding Lieb--Mattis model. This number is extensive (proportional to $N$) since $S$ is extensive. Moreover, the lowest excitation according to Eq.~\eqref{eq:Lieb--Mattis model} has spin $S = \lvert S_A - S_B \rvert \pm 1$ while $S_A$ and $S_B$ are the same, with energy gap $\Delta E = J \left( (s_A - s_B) + \frac{2}{N} \right)$. Crucially, this energy gap is $\mathcal{O}(J)$ instead of order $\mathcal{O}(J/N)$ since $s_A \neq s_B$. 
There is therefore no tower of states with energy gap $\mathcal{O}(1/N)$ that would vanish in the thermodynamic limit. Indeed, the exchange energy $J$ is typically of order 1--10 meV, which is certainly not neglible, possibly even measurable.

Recall that the Lieb--Mattis Hamiltonian is the $\vec{k} =\vec{0},\vec{Q}$ part of the full Heisenberg Hamiltonian. The tower of states consists precisely of the zero-wavenumber excitations, and therefore these states and their energies are identical for both models. We can therefore conclude that the lowest excitations in Heisenberg ferrimagnets will be low (non-zero) wavenumber collective excitations: spin waves with quadratic dispersion whose energy scales as $\mathcal{O}(J/L^2)$ with $L$ the linear system size (so $N = L^d$). While this energy can get arbitrarily low as $L \to \infty$, it will not be as low as the gap in a putative tower of states in $d>2$ dimensions. We will confirm this numerically in Section~\ref{Sec:NumResults}.

In Ref.~\cite{Tasaki:2019}, Tasaki provides a proof of the existence of a tower of states with gaps of order $\mathcal{O}(1/N)$, based on several assumptions. One of these assumptions is that the ground state be unique {\em and} be an eigenstate of a symmetry generator with eigenvalue $M$. In the present case, although the ground states are eigenstates of $\hat{S}^z$, they are degenerate and not unique. Below we will argue that when $M>0$, the ground state is always degenerate. Tasaki's derivation applies therefore only to the case $M=0$ and the results of this section are not in contradiction with the proof. 
The alternative is when the symmetry generator is broken itself in the type-A way, which we shortly discuss in Section~\ref{sec:Outlook}. 

\section{Spontaneous symmetry breaking}
\label{sec:Spontaneous symmetry breaking}
From the standard viewpoint of SSB, the absence of a tower of states naively poses a conundrum. In type-A systems, all $\vec{k}=\vec{0}$ energy eigenstates, including the symmetric ground state, are thermodynamically unstable, and a tiny perturbation will be able to break the symmetry. The existence of a tower of states is necessary to be able to construct the dynamically stable, symmetry-breaking superpositions of energy eigenstates, as the energy fluctuations of these superpositions fall off as $\mathcal{O}(J/N)$. We have just seen the smallest energy gap towards total spin excitations in ferrimagnets is instead $\mathcal{O}(J)$. What does this imply for the symmetry breaking?

The answer is in fact quite simple: the exact ground states already break the symmetry themselves. This is easy to see from the Lieb--Mattis model. We have shown that its spectrum can be assigned definite quantum numbers $S$ and $M_z$ (where the $z$-axis is chosen arbitrarily). The only such state which has full $SU(2)$ spin-rotation symmetry is the one with $S = M_z =0$. (It is not sufficient to have only $M_z=0$. Recall for instance that a two-spin-$\tfrac{1}{2}$ system in the triplet state $s=1,m_z=0$ breaks $S^x$- and $S^y$-rotation symmetry.) The ground states of the ferrimagnet instead have $S = \lvert S_A - S_B \rvert > 0$. In fact, the symmetric state with $S=0$ has quite a high energy compared to this state.

We can also reach this result more formally, by considering the usual SSB procedure of adding an external staggered magnetic field $B$ coupling to the order parameter, to the Lieb--Mattis model:
\begin{equation}
	\hat{H}_{\mathrm{LM}} = \frac{J}{N} \left( \hat{S}^2 - \hat{S}_A^2 - \hat{S}_B^2 \right) 
		- B \left( \hat{S}_A^z - \hat{S}_B^z \right).
	\label{HLM2}
\end{equation}
Here $\hat{S}_{A,B}^z = \sum_{i \in A,B} \hat{S}^z_i$ represent the $z$-component of the total spin on sublattices $A$ and $B$. The matrix elements of the symmetry-breaking field in the basis of Lieb--Mattis eigenstates, $\langle S_A S_B S M_z | ( \hat{S}^z_A - \hat{S}^z_B)  | S_A' S_B' S' M'_z \rangle$, are known exactly (reproduced in Appendix~\ref{Appendix:MatrixElements} for completeness)~\cite{VanWezel:2008aa}. For zero field $B=0$, the ground states have maximal $S_A$, and $S_B$,  minimal $S=|S_A - S_B|$, and are degenerate for any value of $M_z$. For non-zero field, $B>0$, the degeneracy is lifted, and the state with the lowest expectation value of the energy has magnetization $M_z=S$. This state must be a superposition of states with different values of the total spin, $|S_A - S_B| < S < (S_A + S_B)$, because the staggered magnetization operator $S^z_A - S^z_B$ couples state with total spin $S$ to $S\pm 1$ states.

In the limit of large $N$ the matrix elements of the Hamiltonian can be conveniently expressed in terms of the shifted total spin $\tilde{S} = S - |S_A - S_B|$. The Hamiltonian is then, up to order $\mathcal{O}(\frac{1}{N})$:
\begin{align}
  \hat{H} &\approx E_0 + \sum_{\tilde{S},\tilde{S'}}  | S_A S_B \tilde{S} M_z \rangle \left(  f_{\tilde{S}+1} \delta_{\tilde{S}, \tilde{S}'-1} \right. \notag \\
  & \left. ~~~~~~~~~ + a_{\tilde{S}} \delta_{\tilde{S}, \tilde{S}'} + f_{\tilde{S}} \delta_{\tilde{S}, \tilde{S}'+1} \right) \langle S_A S_B \tilde{S}' M_z |, \notag \\
  \text{with}~~~ E_0 &= \frac{J}{4} (s_A - s_B)^2 N + \frac{J}{2} (s_A -s_B) \notag \\ 
    &~~~~~~~~~ - \frac{B}{2} (s_A + s_B) N  - 2 B \frac{s_B}{s_A - s_B}, \notag \\
  a_{\tilde{S}} &= \tilde{S} \left( J (s_A - s_B) + 2 B \frac{s_A + s_B}{s_A -s_B}\right), \notag \\
  f_{\tilde{S}} &= -2B \frac{\sqrt{s_A s_B}}{s_A - s_B} \tilde{S}.
\end{align}
This expression for the Hamiltonian is a special case of a tridiagonal matrix discussed in Appendix~\ref{Appendix:Tridiagonal}. As shown there, the ground state in the large $N$ limit is a superposition of states with different $\tilde{S} = S - |S_A - S_B|$, 
\begin{equation}
	| \psi_0 (B,N) \rangle = \sum_{\tilde{S}} \psi(\tilde{S}) | \tilde{S} \rangle,
\end{equation}
with weights given by
an exponentially decaying function $\psi(\tilde{S}) = c\, \mathrm{e}^{ - \tilde{S} / \lambda}$. The decay `length' $\lambda$ is given by
\begin{align}
  \lambda &= 1 / \log \left( \frac{\frac{1}{2} \left( 1 + \sqrt{1 - 4 \epsilon^2 } \right) }{|\epsilon|} \right), \label{LambdaSol2}\\
  \text{with}~~~ \epsilon &= \frac{ -2 B \sqrt{s_A s_B} }{J (s_A - s_B)^2 + 2 B (s_A + s_B)}.
	\label{EpsSol}
\end{align}
In the limit of zero staggered field $B \rightarrow 0$, the decay length vanishes, and the Lieb--Mattis ground state $| \psi_0 \rangle$ is the total-spin eigenstate with $S = |S_A - S_B|$ and $M_z = S$.

Contrary to the case of type-A SSB systems, the thermodynamic limit and the limit of vanishing field commute for the ferrimagnet:
\begin{equation}
	\lim_{B \downarrow 0} \lim_{N \rightarrow \infty} | \psi_0 (B,N) \rangle 
	= \lim_{N \rightarrow \infty}  \lim_{B \downarrow 0} |  \psi_0 (B,N) \rangle .
\end{equation}
Just as for the ferromagnet, a particular orientation (i.e. the choice of the $z$-axis)  for the ferrimagnetic ground state can be singled out from the ground state manifold by an infinitesimal field even for finite system size. There is thus a discontinuity in the ground state as a function of applied field for any system size:
\begin{equation}
	 \lim_{B \downarrow 0} | \psi_0 (B,N) \rangle  \neq 
	 \lim_{B \uparrow 0}| \psi_0 (B,N) \rangle .
\end{equation}

\begin{figure}[t]
	\includegraphics[width=\columnwidth]{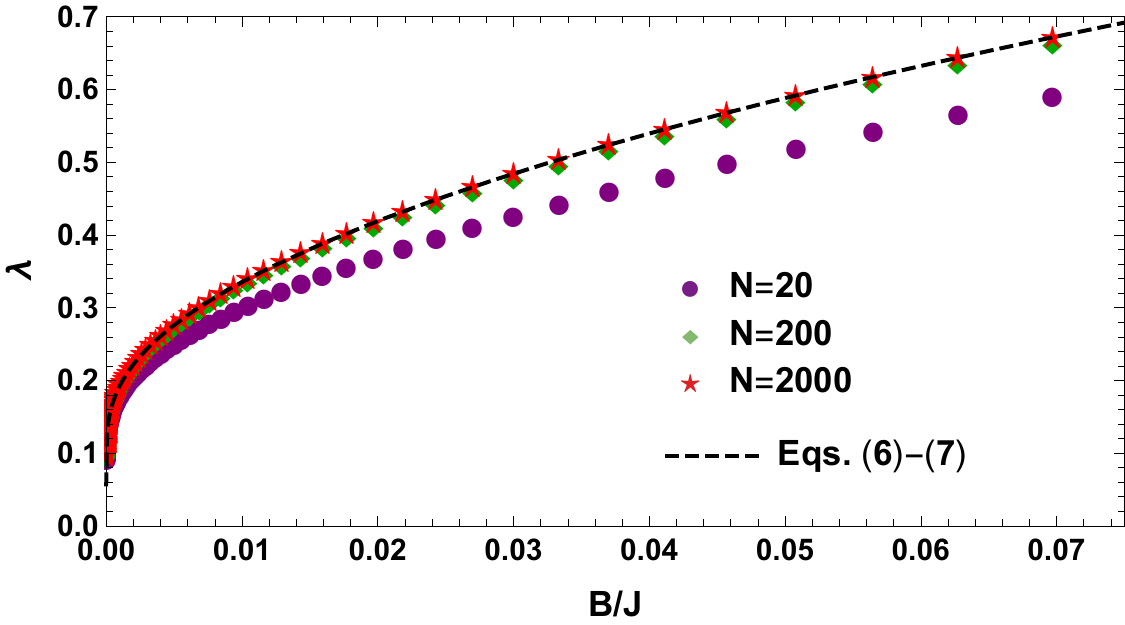}
	\caption{\label{FigLMsol} The ground state of the ferrimagnetic Lieb--Mattis model with staggered magnetic field $B$ is a superposition of states with different $\tilde{S} = S - |S_A - S_B|$, with weights given by $\mathrm{e}^{-\tilde{S}/\lambda}$. Here, a numerically determined decay length $\lambda$ is plotted as a function of $B/J$ for increasing system size up to $N=2000$. The dashed black line is the exact result in the thermodynamic limit following Eqs.~\eqref{LambdaSol2}-\eqref{EpsSol}. It is important to observe that in the thermodynamic limit $N \rightarrow \infty$ the decay length $\lambda$ remains smooth near $B =0$.
	}
\end{figure}

Numerical evaluation of the ground state of the ferrimagnetic Lieb--Mattis Hamiltonian for large but finite $N$ and $B>0$ confirms that it is
an exponentially decaying function in $\tilde{S}$, as shown in Fig. \ref{FigLMsol}. The decay length follows the analytical result of Eqns.~\eqref{LambdaSol2}, \eqref{EpsSol}. 

We conclude that although the ferrimagnet has $2S+1$ degenerate ground states, any small external staggered field is sufficient to lift the degeneracy, upon which the ground state will be remain dominated by the spin state with $S = \lvert S_A - S_B \rvert$, and the magnetization direction parallel to the external field.

\section{The classical N\'{e}el state}\label{sec:Neel}
Besides the total (ferromagnetic) magnetization $M_z$, ferrimagnets also have a nonzero staggered (antiferromagnetic) magnetization $n = \langle \hat{S}^z_A - \hat{S}^z_B \rangle \neq 0$. Using the matrix elements in Appendix~\ref{Appendix:MatrixElements}, the staggered magnetization expectation value for the maximally polarized ground state of the Lieb--Mattis ferrimagnet, with $M_z = S$ and $S = (s_A -s_B) N/2$, and no applied fields, is found to be:
\begin{equation}
  n = N \frac{s_A + s_B}{2} - \frac{2 s_B}{|s_A - s_B|} + \mathcal{O}(N^{-1})
  \label{nLM}
\end{equation}
To leading order, this agrees with the expectation from the classical limit for a ferrimagnet, which is the N\'{e}el product state $| \psi_\textrm{N\'eel}\rangle$ with all spins on the $A$ sublattice pointing maximally up and on the $B$ sublattice maximally down. That state has maximal staggered magnetization $n = N \frac{s_A + s_B}{2}$.

The N\'{e}el state itself can be written as a superposition of states with different total spin $S$ but fixed $M_z = S_A -S_B$, and $S_{A,B} = s_{A,B} N/2$. Because $|\psi_\textrm{N\'eel}\rangle$ is the ground state of the Hamiltonian in Eqn.~\eqref{HLM2} with $J=0$ and $B>0$, the N\'{e}el state wavefunction has the same exponentially decreasing form $\psi(\tilde{S}) = c \mathrm{e}^{ - \tilde{S} / \lambda}$ as the ferrimagnetic ground state. For $J=0$, however, the decay length $\lambda$ in Eqn.~\eqref{LambdaSol2} is independent of $N$ and $B$.
This means that the N\'{e}el state has nonzero overlap with the ground state of the Lieb--Mattis Hamiltonian for any system size, and even in the thermodynamic limit $N \rightarrow \infty$. Its value can be found by normalising the N\'eel state wavefunction:
\begin{equation}
	\langle \psi_\textrm{N\'eel}\; \vert\; S=\lvert S_A -S_B\rvert \rangle = \sqrt{1 - e^{-2 /\lambda(J=0)}}
        \label{LMOverlap}
\end{equation}
For the special case of $s_A = 1$ and $s_B = 1/2$, we have $\epsilon(J=0) = -\frac{\sqrt{2}}{3}$ and the overlap is found to be $\frac{1}{\sqrt{2}}$.

Because the N\'{e}el state is a superposition of total spin states, which do not form a tower of states in the ferrimagnet, the N\'eel state cannot be become energetically degenerate with the ground state of the Lieb--Mattis Hamiltonian for $J>0$ and $B=0$. The energy difference follows directly from
\begin{equation}
	\langle \psi_\textrm{N\'eel} | \hat{H}_\mathrm{LM} | \psi_\textrm{N\'eel} \rangle
	= \frac{J}{4} N |s_A - s_B |^2 + J (s_A^2 + s_B^2).
\end{equation}
This should be compared with the Lieb--Mattis ground state energy $E_0 = \frac{J}{N} (S_A - S_B) (S_A  - S_B + 1) 
=  \frac{J}{4} N |s_A - s_B |^2 + \frac{J}{2} (s_A - s_B)$. The energy difference between $|\psi_\textrm{N\'eel} \rangle$ and the actual ground state is thus:
\begin{equation}
	\Delta E_{N} = J \left(s_A^2 + s_B^2 - \frac{1}{2} s_A  + \frac{1}{2} s_B \right).
\end{equation}
For the specific case with $s_A = 1$ and $s_B = 1/2$, the energy difference is exactly $\Delta E_N = J$. In all cases, it is of order $\mathcal{O}(J)$, and does not vanish in the thermodynamic limit. This is to be contrasted with the antiferromagnet, where the N\'eel state becomes exactly degenerate with the ground state in the thermodynamic limit of the Lieb--Mattis model.

The fact that the staggered magnetization in Eqn.~\eqref{nLM} differs from its classically expected value at sub-leading order, may be interpreted as indicating that the Lieb--Mattis ground state involves zero-wavenumber quantum corrections on top of the classical N\'eel state. These quantum corrections correspond precisely to the suppression of total-spin components outside the ground state manifold. Going towards more realistic models, one should note that the ground state of the Heisenberg ferrimagnet is not the same as that of the Lieb--Mattis Hamiltonian. Because the Heisenberg model includes only local interactions, its ground state will have quantum corrections at all wave numbers as compared to the N\'eel state, and their overlap vanishes in the thermodynamic limit (confirmed numerically below), even though both states will agree to leading order on the expectation value of staggered magnetization. The vanishing overlap is in fact a generic property for ground states of distinct models in an exponentially large Hilbert space, and is expected also for example for the overlap between the ground state of the Heisenberg antiferromagnet and the classical N\'eel state.

\section{Stability}
\label{SecStab}
In Section~\ref{sec:Spontaneous symmetry breaking} we addressed one part of the conundrum that the absence of a tower of states poses, namely how the ground state breaks the symmetry. However, zero-wavenumber energy eigenstates are global---they are not tensor products of local states---and typically unstable as we have shown for instance for the ground state of the antiferromagnet in Section~\ref{sec:Antiferromagnets}. 
The question is whether the symmetry-breaking exact ground states of the finite size ferrimagnet are stable.

In this section we show that the exact ground states of the Lieb-Mattis model with maximal polarization, $M_z = \pm S = \pm \lvert S_A - S_B\rvert$ are thermodynamically stable. In the next section we provide numerical evidence that the maximally polarized ground states in the full Heisenberg model are also stable.

Recall that a state is defined to be {\em unstable} if there exists an extensive observable $\hat{A}$ whose variance $\Var \hat{A} = \langle \hat{A}^2 \rangle - \langle \hat{A} \rangle^2$ scales as $N^2$~\cite{Shimizu:2002PRL,Tasaki:2019,BRvW:lecturenotes}. This definition is equivalent to saying there exists a connected correlation function that does not satisfy the cluster decomposition property~\cite{Shimizu:2002JPSJ,BRvW:lecturenotes}.

To prove that a state is stable one thus needs to compute the variance of all possible extensive observables. In the case of the Lieb--Mattis Hamiltonian in Eqn.~\eqref{HLM2}, all states are global, and we can suffice by computing the variance of the transverse total spin, $\Var \, \hat{S}_x$, and of the total sublattice magnetization, $\Var (\hat{S}^z_A - \hat{S}^z_B)$. 

The variance of the transverse total spin $\hat{S}^x$ is independent of $S_A$ and $S_B$. Because $\langle S M_z | \hat{S}^x | S M_z \rangle = 0$, we find 
\begin{align}
	\Var \left[ \hat{S}^x \right] & = \langle S M_z | (\hat{S}^x)^2 | S M_z \rangle \notag \\
	&= \frac{1}{4} \langle S M_z | (\hat{S}^+ + \hat{S}^-)^2 | S M_z \rangle \notag \\
	&= \frac{1}{4} \langle S M_z | ( \hat{S}^+ \hat{S}^- + \hat{S}^- \hat{S}^+ ) | S M_z \rangle.
\end{align}
The maximally polarized states, with $M_z = \pm S$, are annihilated by $\hat{S}^\pm$, and we then find the variance to be ${S}/{2} \sim \mathcal{O}(N)$.
On the other hand, for $|M_z|< S$ we find:
\begin{equation}
	\Var \left[ \hat{S}^x \right] = \frac{1}{2} S (S+1) - \frac{1}{2} M_z^2.
	\label{VarSxN2}
\end{equation}
As long as $|M_z|$ does not scale with system size in exactly the same way as $S$, the variance of the total transversal spin scales as $\mathcal{O}(N^2)$, implying instability. Therefore only the states with $|M_z|/S \rightarrow 1$ are thermodynamically stable with respect to the total transversal spin.

The variance of the sublattice magnetization $\hat{S}^z_A - \hat{S}^z_B$ can be computed directly using the matrix elements in Appendix~\ref{Appendix:MatrixElements}.
For the maximally polarized ground state with $M_z = S$ and $s_A > s_B$, it is to leading order:
\begin{equation}
	\Var \left[ \hat{S}^z_A - \hat{S}^z_B \right] 
	= 
	4 \frac{s_A s_B}{(s_A - s_B)^2} + \ldots
	\sim \mathcal{O}(1).
        \label{varFiM}
\end{equation}
This shows that the maximally polarized ground state is thermodynamically stable with respect to all extensive observables even in finite-sized ferrimagnets. For states with non-maximal polarization, $M \sim \mathcal{O}(1)$, we have $\Var \left[ \hat{S}^z_A - \hat{S}^z_B \right] = \frac{s_A s_B}{s_A - s_B} N + \ldots$, which also suggests stability. However, since we have already seen that these states with non-maximal polarization are unstable with respect to total transverse spin $\hat{S}^x$, the two maximally polarized ground states are the only thermodynamically stable ground states.

%
\begin{figure}[t]
	\includegraphics[width=\columnwidth]{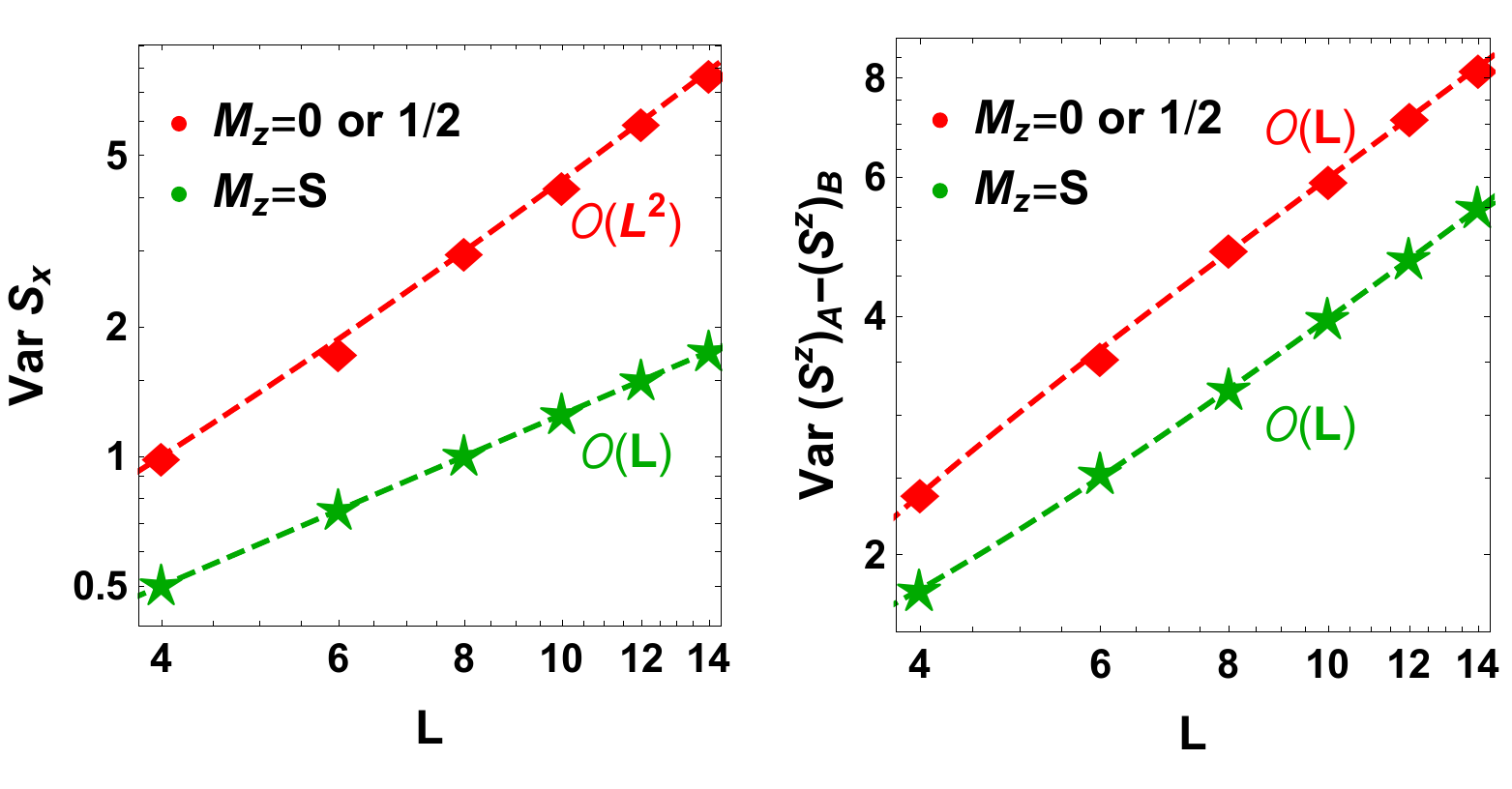}
	\caption{\label{FigVar}
	A numerical comparison of the stabilities of the maximally ($M_z = S = L/2$, green stars) and minimally ($M_z = 0$ or $1/2$, red diamonds) polarized ground states of the 1D ferrimagnetic $s_A = 1$, $s_B = \tfrac{1}{2}$ Heisenberg model.
	The left panel shows $\Var \, \hat{S}^x$, and the results exactly follow Eq.~\eqref{VarSxN2} (shown as dashed lines). The variance in the maximally polarized state scales as $\mathcal{O}(L)$ and thus signal stability, whereas the $\mathcal{O}(L^2)$ scaling in the minimally polarized state implies instability.
	The right panel shows the variance of $\hat{S}^z_A - \hat{S}^z_B$. Here, the results do not match the variance found for the ground states of the Lieb--Mattis model, because $S_A$ and $S_B$ are not symmetries of the ferrimagnetic Heisenberg Hamiltonian. Nevertheless, the variances still scales as $\mathcal{O}(L)$ (the dashed lines represent a fit) which implies that all ground states are stable with respect to the uncertainty of the antiferromagnetic correlations.}
\end{figure}

\begin{figure}[t]
	\includegraphics[width=\columnwidth]{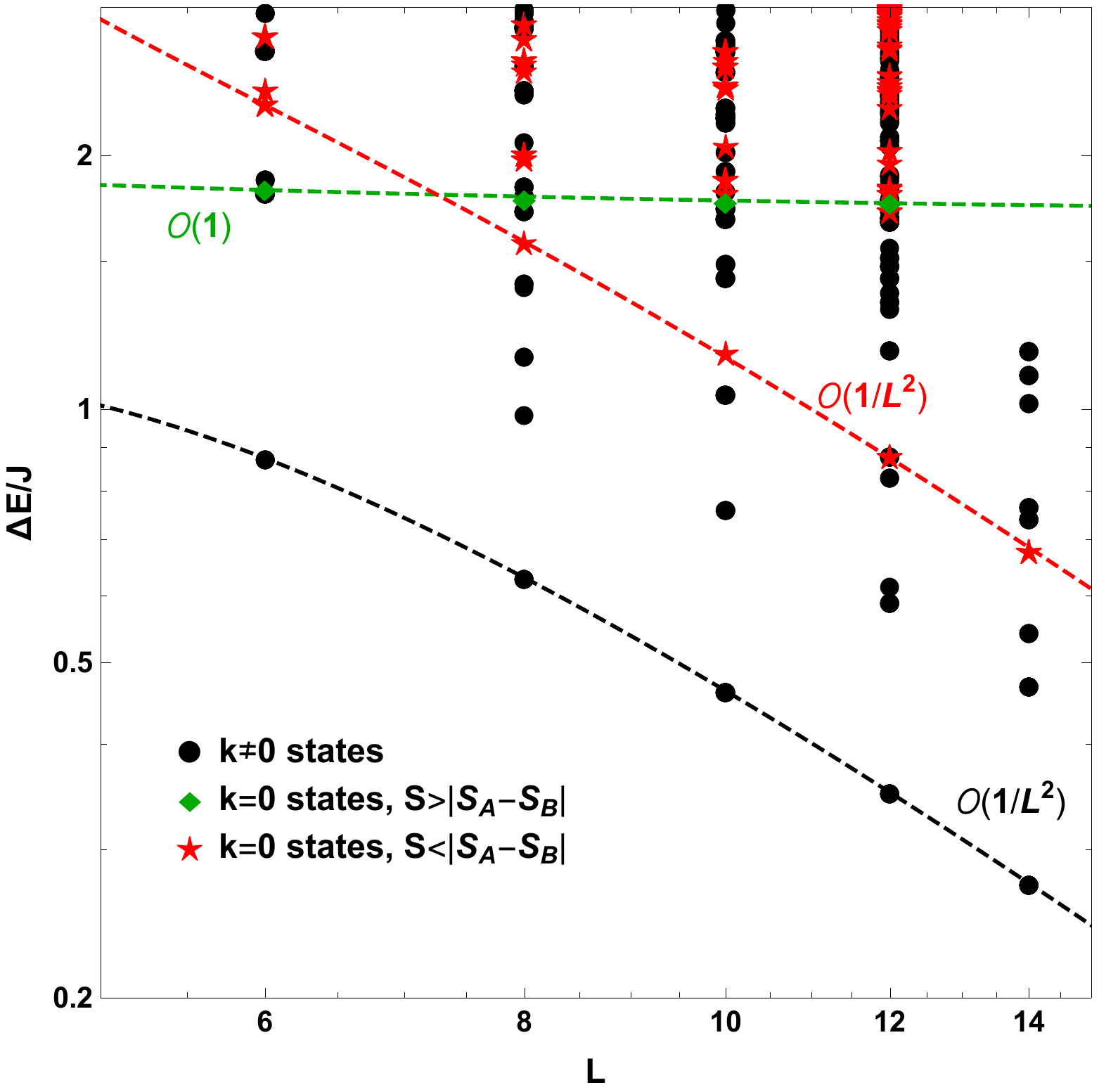}
	\caption{\label{FigGaps}
	An overview of all low-energy excited eigenstates with their energy gap in units of $J$, in the 1D ferrimagnetic $s_A = 1$, $s_B = \tfrac{1}{2}$ Heisenberg chain, as a function of system size $L$. 
	Black dots represent states with finite wave number, $k \neq 0$, while red stars indicate states with zero wave number $k=0$ and $S < |S_A - S_B|$, and green diamonds have $k=0$ and $S > |S_A - S_B|$. Here $|S_A - S_B| = L/4$ is the ground state total spin. 
	In one dimension, the excitation gap towards Goldstone modes with nonzero wave number (black dots) is $\mathcal{O}(1/L^2)$ (indicated by the black dashed line),  because $\epsilon_k \sim k^2$ and the smallest momentum scales as $k\sim \mathcal{O}(1/L)$. 
	States with $k=0$ and $S<|S_A - S_B|$ are two-Goldstone mode states (red stars), which follows from the fact that their magnetization eigenvalue $M_z$ is smaller than the ground state value. 
	Finally, the states that would constitute the tower of states in an antiferromagnet, namely those with $k=0$ and $S>|S_A - S_B|$ (green diamonds), have a gap that is independent of system size (green dashed line).
	}
\end{figure}

\section{Numerical results}
\label{Sec:NumResults}
The Mermin--Wagner--Hohenberg--Coleman theorem does not prohibit type-B SSB from occurring in one-dimensional systems at zero temperature~\cite{Watanabe:2014bg,Griffin:2015,BRvW:lecturenotes}. We can therefore confirm the generality of the analytic results of the Lieb--Mattis model in Section~\ref{SecStab} using numerical results for a one-dimensional ferrimagnetic Heisenberg chain. The Hamiltonian is given by:
\begin{equation}
	\hat{H} = J \sum_i \hat{\vec{S}}_i \cdot \hat{\vec{S}}_{i+1}.
\end{equation}
For concreteness, we take all the even sites to have $s_A=1$ spins and the odd sites $s_B=1/2$ spins. We consider chains up to lengths $L=14$ using exact diagonalization, and evaluate degeneracies and the stability of the ground states. Also note that because our system is one-dimensional, the linear size $L$ and total system size $N$ are equal.

We confirm that the $s_A = 1$, $s_B = \tfrac{1}{2}$ Heisenberg ferrimagnet has a $L/4+1$-dimensional ground state manifold with total spin $S=L/4$, just like the Lieb--Mattis ground states. 

Next we analyze the stability of the states in the ground state manifold. Following Sec.~\ref{SecStab}, we compute the variance of the transverse total spin and the staggered magnetization, shown in Fig.~\ref{FigVar}. The variance of staggered magnetization scales with the system size, independent of the magnetic number $M_z$. All states in the ground state manifold are therefore stable with respect to staggered magnetization. However,  the variance of transverse spin scales as $\mathcal{O}(L^2)$ for the state with minimal $M_z$ but it scales as $\mathcal{O}(L)$ for the maximally polarized state $M_z = S$. In fact, the results are exactly equal to Eq.~\eqref{VarSxN2}. We therefore confirm that the only thermodynamically stable ground state is the maximally polarized state with $M_z =S$.

The absence of a tower of states is confirmed by an analysis of the low-lying eigenstates, shown in Fig.~\ref{FigGaps}. For each low-energy eigenstate we computed its energy, momentum $k$, total spin $S$ and polarization $M_z$. If a tower of states would be present, the energy gap towards states with $k=0$, $M_z = |S_A - S_B|$ and $S > |S_A - S_B|$ would vanish as $\mathcal{O}(1/L)$. This is not the case, as we argued in Sec.~\ref{SecToS}: in Fig.~\ref{FigGaps} we see that these states (green diamonds) have a gap of order $\mathcal{O}(1)$ (green dashed line).

Nonetheless, there are states whose gap vanishes as $\mathcal{O}(1/L^2)$. We identify here the excitation of a Goldstone mode, with dispersion $\epsilon \sim k^2$ and the smallest momentum scaling as $k \sim \mathcal{O}(1/L)$. Goldstone modes necessarily have nonzero momentum, shown in Fig.~\ref{FigGaps} as black circles. The excitation of two Goldstone modes with opposite momentum now leads to states with zero momentum but still a gap that scales as $\mathcal{O}(1/L^2)$, shown as red stars. We thus have analyzed the full eigenvalue spectrum and excluded the possibility of a tower of states.

\begin{figure}
	\includegraphics[width=\columnwidth]{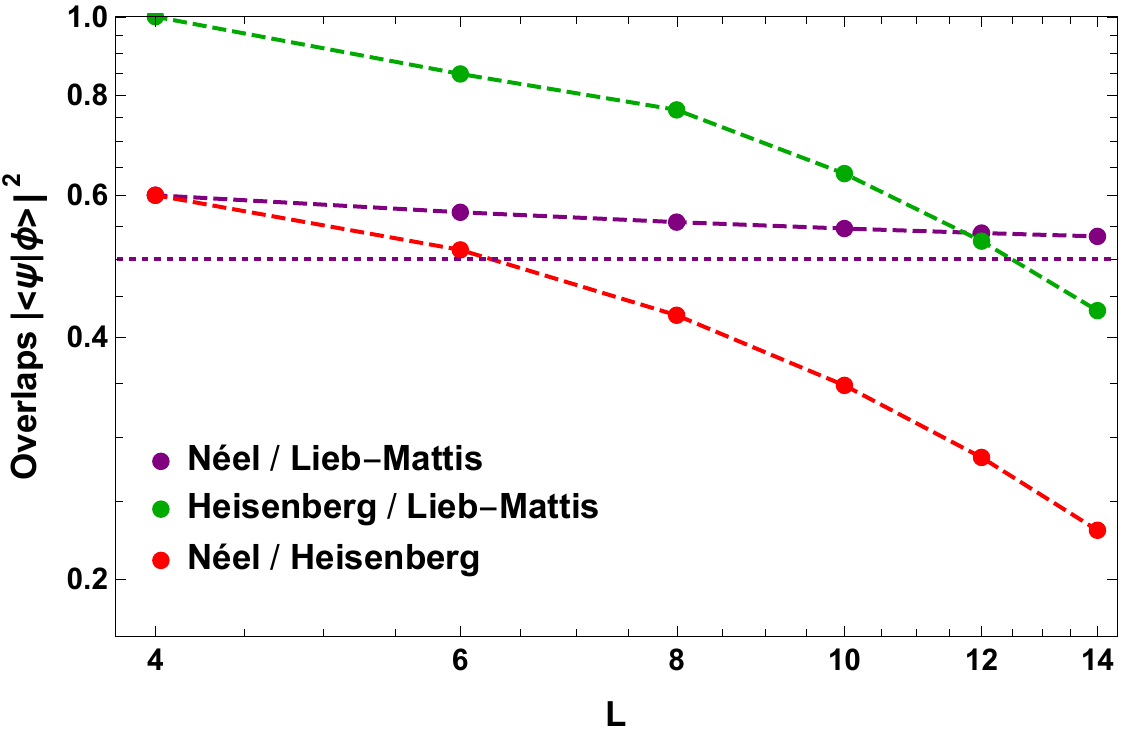}
	\caption{\label{FigOverlaps}The overlap-squared between ground states of various models shown on a log-log scale.
	The overlap between the maximally polarized ground state of the Lieb--Mattis model state and the classical N\'{e}el state follows Eqn.~\eqref{LMOverlap} and stays nonzero even in the thermodynamic limit.
	On the other hand, the presence of spin flips or quantum corrections in the maximally polarized ground state of the ferrimagnetic Heisenberg model causes its overlaps with both the N\'{e}el state and the Lieb--Mattis state to vanish with increasing system size.}
    \end{figure}

Finally, Fig.~\ref{FigOverlaps} shows the overlap of the  Lieb--Mattis and Heisenberg ground states with the classical N\'{e}el state. While the former have a nonzero overlap in the thermodynamic limit, the overlap of the latter decreases with system size. This can be understood as a consequence of the extensive number of single and few-spin flips contained in the Heisenberg ground state relative to the N\'{e}el state. These can be seen as quantum corrections to the classical state at all wave numbers, and although they hardly affect the macroscopic staggered magnetization, and do not affect the magnetization at all, they do cause the overlap with the N\'eel state to vanish in the thermodynamic limit. The latter also occurs in the (type-A) antiferromagnet.

\section{Outlook}\label{sec:Outlook}

We found the ferrimagnet to have an extensive ground state degeneracy and no tower of states. Furthermore, we found that within the ground state manifold, the maximally polarized states are thermodynamically stable. The ferrimagnet shares these features with the ferromagnet, which turns out to be less unique than often assumed~\cite{Watanabe:2012jn,Watanabe:2014bg,Beekman:2015gr}.

Although we found these results for the specific case of the ferrimagnet, we hypothesize that these conclusions apply to type-B systems in general. The defining property of such systems is 
that the the expectation value of the commutator of two broken symmetry generators does not vanish. Apart from some pathological cases~\cite{Brauner:2010review}, such a commutator is a linear combination of symmetry generators itself, implying that {\em the order parameter operator commutes with the Hamiltonian}, i.e. the order parameter operator is a symmetry generator.

Let us consider the Lie algebra structure of the Hamiltonian and its eigenstates. The symmetry generators $\hat{Q}^a$, which by definition are Hermitian and commute with the Hamiltonian, can be expressed in the Cartan--Weyl basis, in which the Cartan subalgebra is spanned by a maximal set of $r$ mutually commuting generators $\hat{F}^i$, where $r$ is called the rank of the Lie algebra. The remaining generators can be expressed in pairs of Hermitian-conjugate root generators $\hat{E}^\alpha, \hat{E}^{-\alpha}$, with $\alpha$ called the root vector and commutation relations $[\hat{E}^\alpha , \hat{E}^{-\alpha}] = \alpha_i \hat{F}^i$. Watanabe and Brauner have shown that the Cartan subalgebra can be chosen in such a way that only the Cartan generators $\hat{F}^i$ can obtain an expectation value, and hence lead to type-B SSB~\cite{WatanabeBrauner:2011}. We can simultaneously diagonalize the Hamiltonian and the Cartan generators. Eigenstates of the Cartan generators $\hat{F}^i$ are {\em weight states} $\lvert \mu \rangle$ with eigenvalues $\mu_i$, collected in a weight vector $\mu$.

Now we recall several important theorems from Lie group theory~\cite{georgi2018lie,iachello2006lie}. First, irreducible representations of a semisimple Lie algebra are completely classified by specifying the highest weight $\bar{\mu}$.  Second, a Lie algebra of rank $r$ contains $r$ Casimir operators, which commute with the entire Lie algebra. Third, by Schur's lemma, any operator that commutes with all generators of the Lie algebra is proportional to the unit matrix in any irreducible representation.

For discussing the ground states and the tower of states, we only need to consider the $k=0$-part of the Hamiltonian, which therefore only depends on internal degrees of freedom which transform under the Lie group of symmetry transformations. Consequently we are going to assume that the $k=0$  Hamiltonian can be completely specified in terms of Lie algebra generators, in other words it is a {\em spectrum-generating algebra}~\cite{iachello2006lie}. It follows that the $k=0$  Hamiltonian, which commutes with the entire algebra, consists only of Casimir operators.

For type-B SSB, a ground state is an eigenstate of symmetry generators in the Cartan subalgebra, i.e. a weight state, where at least one weight component is non-zero. Since a symmetry generator is an extensive operator, this weight component is extensive. Because the irreducible representation of this generator is specified by its highest weight, the highest weight must also be extensive, which implies that the representation space has extensive dimensions. And since the Hamiltonian consists of Casimir operators, which are proportional to the identity matrix in any irreducible representation, this implies that there is a ground state degeneracy of the extensive dimension of this representation. This proves that under mild assumptions type-B SSB involves an extensive ground state degeneracy.

As for the tower of states, we shall discuss the case where the $k=0$ Hamiltonian is proportional to the quadratic Casimir operator $\hat{C}_2$, which consists of quadratic combinations of generators $\hat{Q}^a$. This comprises most models of interest, including the Heisenberg ferrimagnet and (anti)ferromagnet. Because the $\hat{Q}^a$ are extensive, the Hamiltonian must be of the form $\hat{H}_{k=0} = \frac{1}{N} \hat{C}_2$, up to factors of order $\mathcal{O}(1)$, so that the energy is extensive. In the Cartan-Weyl basis, the quadratic Casimir operator can be expressed as~\cite{Ma:2007}
\begin{equation}
 \hat{C}_2 = \sum_i \hat{F}^i \hat{F}^i + \sum_{\alpha \in \Delta_+} (\hat{E}^\alpha \hat{E}^{-\alpha} + \hat{E}^{-\alpha} E^{\alpha}),
\end{equation}
where $\Delta_+$ is the set of positive root vectors. Since Casimir operators are proportional to the identity matrix, we can find the proportionality constant by acting on the highest weight state for which $\hat{E}^\alpha \lvert \bar{\mu}\rangle =0$. Then, acting on this state, the term in brackets is equal to $[\hat{E}^\alpha,\hat{E}^{-\alpha}] = \alpha_i \hat{F}^i$. We therefore find the energy of all states in the irreducible representation $\bar{\mu}$ to be
\begin{equation}\label{eq:Casimir Hamiltonian}
 E = \frac{1}{N} \sum_i \bar{\mu}_i \big(\bar{\mu}_i + \sum_{\alpha \in \Delta_+} \alpha_i\big).
\end{equation}

With all this, we can analyze the putative tower of states. Denoting the representation to which the ground state belongs by $\bar{\mu}_0$, it consists of $k=0$ states in different irreducible representations $\bar{\mu}'$, but with the same eigenvalues $\mu$ for the Cartan generators, which implies that $\bar{\mu}' > \bar{\mu}_0 \ge \mu$. From Eq.~\eqref{eq:Casimir Hamiltonian} we see that, since $\bar{\mu}'$ is extensively non-zero, the energies of any excited state contain at least a factor of $\frac{1}{N}\bar{\mu}'_j$, and therefore are at least $\mathcal{O}(1)$. There is no tower of states with gaps $\mathcal{O}(1/N)$.

There is one caveat: it could be that the weight state has eigenvalue $\mu_j = 0$ for one (or more, but not all) Cartan generators. Then Eq.~\eqref{eq:Casimir Hamiltonian} would allow for a tower of states with energy gaps $\mathcal{O}(1/N)$. But if a Cartan generator has eigenvalue 0, there are two possibilities: first, the state could be invariant under one or more $SU(2)$ subgroups, in which case the root generators which would construct the tower of states also annihilate the state. Second, the symmetry is broken in the type-A way, in which case a tower of states is expected. We leave detailed investigation of systems with both type-A and type-B breaking for future work.

We therefore conclude that states which feature type-B SSB exclusively, have an extensive ground state degeneracy and no tower of states with energy gaps $\mathcal{O}(1/N)$.

The distinction between symmetry breaking where the expectation value of the Casimir operator is minimal or maximal has been discussed before~\cite{Higashikawa:2016dq}, although these authors do not consider the case where $\mu < \bar{\mu}$ as is the case for almost all type-B SSB~\cite{Beekman:2015gr}.

The nonzero overlap between the ground state of the ferrimagnetic Lieb--Mattis Hamiltonian and the classical N\'{e}el state suggests it may be possible to find a simple exact representation of this ground state, which we leave for future work. It also opens up the question of the entanglement structure of the ferrimagnet, which unlike that of the product state Heisenberg ferromagnet, is quite subtle. It would be interesting to see whether the entanglement in type-B SSB systems exhibits the same Goldstone mode counting as type-A SSB systems.\cite{Metlitski:akSbyvLD,Rademaker:2015ie}

\begin{table}[t]
 \begin{tabular}{cccccc}
 \toprule
 SSB & \thead{\# ground \\states} & \thead{tower of\\ states} & \thead{\# NG \\ modes } & \thead{NG \\ dispersion} & \thead{lower critical \\dimension} \\
 \cmidrule(r){1-1} 
\cmidrule(l){2-6}
  type-A & 1 & yes & $n$ &linear & 1 \\ 
  type-B & $\mathcal{O}(N)$ & no & $\tfrac{n}{2}$ & quadratic & 0\\
  \bottomrule
 \end{tabular}
\caption{Comparison between type-A and type-B SSB phenomenology. They differ in  the number of ground states; the existence of a tower of states; the number of Nambu--Goldstone modes in terms of the number of broken symmetry generators $n$; their dispersion relation; and the lower critical dimension, which is the lowest dimension at which zero-temperature SSB is possible (the lower critical dimension at finite temperature is 2 for both type-A and type-B SSB).
}\label{table:overview}
\end{table}

Concluding, the distinction between type-A and type-B SSB seems to go much further than the counting of Goldstone modes~\cite{Watanabe:2012jn,Hidaka:2013}, as is summarized in Table~\ref{table:overview}. Type-A, ordinary, SSB has a unique symmetric ground state and a tower of low-lying states with energy gap $\sim \mathcal{O}(1/N)$. There is a linearly dispersing Goldstone mode for each broken symmetry. Furthermore it has quantum corrections to the classical SSB state, and due to the Coleman theorem type-A SSB in one dimension at zero temperature is forbidden in the thermodynamic limit. Conversely, here we have found that type-B SSB is accompanied by an extensive ground state degeneracy and has no tower of low-lying states. Instead, at least one of the ground states is thermodynamically stable. Two broken symmetry generators lead to one quadratically dispersing Goldstone mode and a gapped partner mode. Finally, type-B systems do not suffer from the Coleman theorem and are stable in one dimension~\cite{Watanabe:2014bg,Griffin:2015,BRvW:lecturenotes} (although both type-A and type-B systems are subject to the Mermin--Wagner--Hohenberg theorem that forbids SSB in two or lower dimensions at {\em finite} temperature~\cite{Watanabe:2014bg,BRvW:lecturenotes}).
There seems to be only one essential difference between general (``ferri'') type-B SSB and the peculiar case of the ferromagnet: 
the latter is the same as the classical ferromagnet, whereas the ferrimagnet is a classical N\'{e}el state dressed with quantum corrections~\cite{Beekman:2015gr}. 
Another difference is that the ferromagnet does not contain a gapped mode partnered with the gapless  Goldstone mode; this can be interpreted as the gap being pushed to infinity~\cite{Kobayashi:2015,BRvW:lecturenotes}.
Nevertheless, here we have seen that even if quantum corrections are present in type-B systems, the lowest energy gap of zero-wavenumber states is not $\mathcal{O}(1/N)$ but $\mathcal{O}(1)$ and they do not constitute a tower of states in the sense of Refs.~\cite{Anderson:2011vu,Horsch:1988,Kaiser:1989,Kaplan:1990,Koma:1994,Tasaki:2019}.

\acknowledgments
We thank Hal Tasaki for inspiring discussions and for sharing a draft version of Ref.~\cite{Tasaki:2019book}.
This work is supported by the Swiss National Science Foundation via an Ambizione grant (L.~R.); by the MEXT-Supported Program for the Strategic Research Foundation at Private Universities ``Topological Science'' (Grant No. S1511006) and by JSPS Grant-in-Aid for Early-Career Scientists (Grant No. 18K13502) (A.~J.~B.);

\appendix

\section{Matrix elements}
\label{Appendix:MatrixElements}
For completeness, we reproduce here the matrix elements of the staggered magnetic field in the basis of Lieb--Mattis eigenstates $| S_A S_B S M_z \rangle$, as described in Ref.~[\onlinecite{VanWezel:2008aa}]:
\begin{align}
	&\langle S_A S_B S M_z | ( S^z_A - S^z_B)  | S_A' S_B' S' M'_z \rangle
	=\delta_{S_A, S_A'} \delta_{S_B, S_B'} 
	\notag \\
	&~~~\times \delta_{M_z, M'_z} 
	\left[ f_{S+1} \delta_{S, S'-1} + g_{S} \delta_{S, S'} + f_{S} \delta_{S, S'+1} \right], 
	\label{Me} 
\end{align}
Here, we defined the functions:
\begin{eqnarray}
	f_S  \equiv  {\scriptstyle \sqrt{ \frac{(S^2 - (S_A - S_B)^2) ((S_A + S_B + 1)^2 - S^2) (S^2 - M_z^2)}{(2S+1)(2S-1)S^2} }}, \notag \\
	g_S  \equiv  \frac{(S_A - S_B) (S_A + S_B + 1) M_z }{S(S+1)}.
	\label{Me3} 
\end{eqnarray}
For the specific case of $S_{A/B} = s_{A/B} N /2$, $M_z=0$, and large $N$, the matrix elements can be conveniently expressed in terms of the shifted total spin $\tilde{S} = S - |S_A - S_B|$, up to order $\mathcal{O}(1/N)$:
\begin{align}
	f_{\tilde{S}} & \approx 2 \frac{\sqrt{s_A s_B}}{s_A - s_B} \tilde{S} \notag \\
	g_{\tilde{S}} & \approx \frac{1}{2} (s_A + s_B) N - 
		2 \frac{ s_B + (s_A + s_B) \tilde{S} }{s_A - s_B} \notag  \\
	S(S+1) & \approx
                           \frac{1}{4} (s_A - s_B)^2 N^2 \notag \\
  &~~~~~~~ +
		\frac{1}{2} (s_A - s_B) (2 \tilde{S}+1) N
\end{align}

\section{Ground state of a tridiagonal matrix}
\label{Appendix:Tridiagonal}
Consider a symmetric tridiagonal matrix, where the only nonzero elements are on the diagonal and just below and above it:
\begin{align}
	M= \begin{pmatrix} 
	a_1 & b_1 &  &  & \\
	b_1 & a_2 & b_2 & & \\
	 & b_2 & a_3 & b_3 & \\
	 &       & \ddots & \ddots & \ddots 
	\end{pmatrix}.
\end{align}
We are interested in the ground state eigenvector and eigenvalue for the special case where both $a_x$ and $b_x$ increase linearly with $x>0$. By rescaling the matrix we define:
\begin{equation}
	a_x = x, \; \; b_x = \epsilon x.
\end{equation}
      
As an ansatz for the ground state eigenvector, with eigenvalue $v$, we choose:
\begin{equation}
	\psi_x = (-\mathrm{sgn} [\epsilon])^x e^{-x/\lambda}.
\end{equation}
The eigenvalue equation now reads, for each row $x$:
\begin{equation}
	| \epsilon | \left( (x-1) e^{1/\lambda} +  x e^{-1/\lambda} \right) =
	x - v.
\end{equation}
By looking at the first row, with $x=1$, we can relate $\epsilon$, $\lambda$ and $v$:
\begin{equation}
	|\epsilon| e^{-1/\lambda} = 1 - v.
\end{equation}
Inserting this back into the equation for general $x>1$, we find:
\begin{equation}
	|\epsilon|^2 \frac{x-1}{1-v} + x (1-v) = x - v,
\end{equation}
which can be simplified to:
\begin{equation}
	(x-1) (|\epsilon|^2 - v (1-v)) = 0.
\end{equation}
Because the second factor must equal zero for all $x$, we found an expression for $v$ in terms of $\epsilon$ that is independent of $x$. This proves that $\psi_x$ is indeed an eigenvector of $M$ with eigenvalue:
\begin{equation}
	v = \frac{1}{2} \left( 1 + \sqrt{1 - 4 |\epsilon|^2 } \right).
\end{equation}
The exponential decay length is given by:
\begin{equation}
	\lambda = 1 / \log \left( \frac{\frac{1}{2} \left( 1 + \sqrt{1 - 4 |\epsilon|^2 } \right) }{|\epsilon|} \right).
\end{equation}


%

\end{document}